\documentclass[pra,twocolumn,superscriptaddress,showpacs]{revtex4}
\usepackage{amsfonts,amssymb,amsbsy,bm,graphicx,times,color}

\newcommand{\Tr}{\mathop{\mathrm{tr}}\nolimits}
\newcommand{\op}[1]{\hat{#1}}

\begin{document}

\title{Angular performance measure for tighter uncertainty relations}

\author{Z. Hradil}

\affiliation{Department of Optics, Palacky University, 17.
listopadu 50, 772 00 Olomouc, Czech Republic}

\author{J. \v{R}eh\'{a}\v{c}ek}
\affiliation{Department of Optics, Palacky University, 17.
listopadu 50, 772 00 Olomouc, Czech Republic}

\author{A. B. Klimov}
\affiliation{Departamento de F\'{\i}sica, 
Universidad de Guadalajara, 44420~Guadalajara, 
Jalisco, Mexico}

\author{I. Rigas}
\affiliation{Departamento de \'Optica, Facultad de F\'{\i}sica,
Universidad Complutense, 28040~Madrid, Spain}

\author{L. L. S\'{a}nchez-Soto}
\affiliation{Departamento de \'Optica, Facultad de F\'{\i}sica,
Universidad Complutense, 28040~Madrid, Spain}

\begin{abstract}
  The uncertainty principle places a fundamental limit on the accuracy
  with which we can measure conjugate quantities. However, the
  fluctuations of these variables can be assessed in terms of
  different estimators. We propose a new angular performance that
  allows for tighter uncertainty relations for angle and angular
  momentum. The differences with previous bounds can be significant
  for particular states and indeed may be amenable to experimental
  measurement with the present technology.
\end{abstract}


\maketitle

Apart from interpretational issues, the main goal of quantum mechanics
is to make predictions on the outcomes of experiments.  In fact, in many
modern setups one is led to measurements that simultaneously estimate
two  noncommuting variables. The precision with which they are jointly
estimated  obey a fundamental constraint dictated by the uncertainty
principle~\cite{Peres:1993pj}.

The archetypal example is the case of continuous variables, such as
position and linear momentum of a single particle. The standard
formalization of the uncertainty principle is presented in terms of
the associated variances [defined as $(\Delta A)^{2} = \langle
\op{A}^{2} \rangle - \langle \op{A} \rangle$], and it
reads~\cite{Heisenberg:1967gu} (with $\hbar = 1$ throughout)
\begin{equation}
  \label{upxp}
  (\Delta x)^2 \, (\Delta p)^2 \ge \frac{1}{4} \, .
\end{equation}
These variances are a measure of the width of the corresponding
probability distributions in the quantum state. However, it has long
been argued that some experiments do not measure variances and
encouraging reformulations of Eq.~(\ref{upxp}) have been proposed in
terms of other resolution
measures~\cite{Heinonen:2007nm,Schurmann:2009mo}.  In other words, one
can assign different measures of inaccuracy (each one with its own
pros and cons) to a particular measurement and this proves crucial to
properly set its ultimate resolution limits.  The price one has to pay
is that establishing an uncertainty principle in terms of these
measures can turn out to be very
intricate~\cite{Wootters:1979wv,Englert:1996tn,Durr:2001by,Hall:2004yt,Ozawa:2004uq}.
The situation is even more ambiguous for magnitudes that cannot be
measured, but must be only inferred, as it happens with, e. g.,
entanglement~\cite{Horodecki:2009hm}.

Angular variables are also riddled with the same kind
of problems, but aggravated by the peculiarities of their periodic
character~\cite{Lynch:1995rl,Perinova:1998pc,Luis:1998al,Luis:2000po}.
Though this is an old question, it experiences periodic revivals
in connection with some hot topics. Nowadays, a renewed interest
in these features has been triggered by the treatment of rotating
Bose-Einstein condensates~\cite{Leggett:2001fs,Cornell:2002sb} and
the quantum optics of vortex beams~\cite{Allen:2003}.  It is worth
remarking that we have at hand very simple experimental schemes 
to test in practice ideal angle concepts.

There is agreement in using the variance $(\Delta L)^2$ to
characterize fluctuations in angular momentum (although, since this
variable is unbounded, the variance may fail in some instances to
provide a satisfactory expression for the uncertainty
principle~\cite{Uffink:1990}).  In contrast, there is no wide
consensus concerning the proper assessment of the conjugate angle
fluctuations.  Periodicity may lead to serious troubles when using 
variance, since the powers of the angle are not periodic functions, so 
that their mean values depend on the origin chosen. There are several 
proposals that avoid these problems, such as the S\"{u}ssmann
measure~\cite{Birula:1993tw,Hall:1999os,Luis:2006il}, circular
variance~\cite{Levy:1976pj,Breitenberger:1985uf,Bandilla:1991ee,Hradil:1992cq,Opatrny:1994to},
entropies~\cite{Birula:1975kx,Deutsch:1983pq,Maassen:1988ta,Abe:1992pt,Brukner:2001kc},
reciprocal peak
height~\cite{Shapiro:1991zv,Schleich:1991hc,Hradil:1992yb},
origin-optimized angle variance~\cite{Trifonov:2003lo,Pegg:1997mm},
and other nonstandard
quantities~\cite{Landau:1961ns,Kowalski:2002}. In short, for periodic
variables there are a lot of candidates for assessing fluctuations,
each one surely with its virtues, but no undisputed champion.

As commented before, if we decide to choose, e. g., the circular
variance (which is computed as the standard one, but using the moments
of the complex exponential of the angle rather than the angle itself
and is the simplest natural choice from a pure statistical
viewpoint~\cite{Mardia:2000}), the resulting uncertainty relation is
rather involved and cannot be saturated, except in very trivial
cases~\cite{Hradil:2006pl,Rehacek:2008ss}.

All these difficulties motivate this paper. We shall seek for
a new angular performance measure that, apart from properly
quantifying angle fluctuations, provides simple and feasible
bounds for the conjugate variable.

To be as self-contained as possible, we first introduce some basic
notions for the problem at hand. We are concerned (assuming cylindrical
symmetry) with the planar rotations by an angle $\phi$ generated by
the angular momentum along the $z$ axis, which for simplicity will be
denoted henceforth as $\op{L}$. Classically, a point particle is
necessarily located at a single value of the periodic angular
coordinate $\phi$, defined within a chosen window. The corresponding
quantum wave function, however, is an object extended around the unit
circle and so can be directly affected by the nontrivial topology.

One may be tempted to think that angle should stand in the same
relationship to angular momentum as ordinary position stands to linear
momentum. This would prompt to interpret the angle operator as 
multiplication by $\phi$ while $\op{L}$ is the differential
operator $\op{L} = - i \partial_{\phi}$. However, the use of this
operator may entail many pitfalls for the unwary: in particular,
single-valuedness restricts the Hilbert space to the subspace of
$2\pi$-periodic functions, which, among other things, rules out the
angle coordinate as a \textit{bona fide}
observable~\cite{Carruthers:1968,Emch:1972}.

Many of these difficulties can be avoided by simply selecting
angular coordinates that are both periodic and continuous instead.
A single such quantity cannot uniquely specify a point on
the circle because periodicity implies extrema, which excludes a
one-to-one correspondence and hence is incompatible with
uniqueness. Perhaps the simplest
choice~\cite{Louisell:1963,Mackey:1963} is to adopt two angular
coordinates, such as, e. g., cosine and sine. In classical
mechanics this is indeed of a good definition, while in quantum
mechanics one would have to show that these variables, we shall
denote by $\op{C}$ and $\op{S}$ to make no further assumptions
about the angle itself, form a complete set of commuting
operators.  One can concisely condense all this information using
the complex exponential of the angle $\op{E} = \op{C} - i \op{S}$,
which satisfies the commutation relation
\begin{equation}
  \label{ELE}
  [ \op{E},  \op{L} ] = \op{E} \, .
\end{equation}
In mathematical terms, this defines the Lie algebra
of the two-dimensional Euclidean group E(2). Interestingly
enough, E(2) is the canonical symmetry of the cylinder, which
is the phase space for our system.

The action of $\op{E}$ on the angular momentum basis $| \ell
\rangle$ is $\op{E} | \ell \rangle = | \ell -1 \rangle$, and it
possesses then a simple implementation by means of phase mask removing
a unit charge from a vortex state~\cite{Hradil:2006pl}.  Since the
integer $\ell$ runs from $- \infty$ to $+ \infty$, $\op{E}$ is a
unitary operator whose eigenvectors
\begin{equation}
  \label{fi}
  |\phi \rangle = \frac{1}{\sqrt{2 \pi}} \sum_{\ell \in \mathbb{Z}}
  e^{i \ell \phi} \, |\ell \rangle 
\end{equation}
describe states with well-defined angle. Although the proposal that
this operator represents the angle conflicts with the orthodox view of
describing observables by Hermitian operators, the option for $\op{E}$
is actually very natural.  Note that one could expect a Fourier
relationship between angle and angular momentum. In this
context, this can be expressed as
\begin{equation}
  e^{- i \phi^\prime \hat L}
  | \phi \rangle = | \phi - \phi^{\prime} \rangle \, ,
\end{equation}
which can be easily verified by using the explicit form in 
Eq.~(\ref{fi}).

Let us turn to the corresponding uncertainty relations. The Robertson
inequality~\cite{Robertson:1929ly,Sudarshan:1995ty} (which remains 
valid for unitary operators) can be applied to obtain
\begin{equation}
  \label{disp}
  \ (\Delta L )^2  \ge  \frac{1}{4} 
   \frac{[ 1 - ( \Delta E)^2]}{(\Delta E)^{2}}   \, ,
\end{equation}
where we have rearranged terms to facilitate comparison with the next
steps in our analysis. Here we have used the natural extension of
variance for unitary operators~\cite{Levy:1976pj}
\begin{equation}
  \label{defDE}
  (\Delta E )^2 = \langle \op{E}^\dagger \op{E} \rangle -
  \langle \op{E}^\dagger \rangle \langle \op{E} \rangle =
  1 - | \langle \op{E} \rangle |^2 \, ,
\end{equation}
which it exactly agrees with the circular variance~\cite{Mardia:2000}.  
The form (\ref{disp}) has been advocated by many authors.  However, 
although correct, it does not provide the tightest lower bound and 
equality cannot  be attained except for some trivial states~\cite{Kowalski:2002}.

To face this disadvantage, let us first recast Eq.~(\ref{ELE})
in terms of the corresponding Hermitian components
\begin{equation}
  \label{CLS}
 [\op{C}, \op{L} ] =  i \op{S} ,
 \qquad
 [\op{S},  \op{L} ] = -  i \op{C} \, ,
\end{equation}
while $[\op{C}, \op{S} ] = 0$. Moreover, for reasons that will
be apparent soon, we look at their rotated versions
\begin{equation}
  \label{rotation}
  \op{C}_{\alpha} = \op{C} \cos \alpha - \op{S}  \sin \alpha \, ,
  \qquad
  \op{S}_{\alpha} = \op{S} \cos \alpha + \op{C}  \sin \alpha \, .
\end{equation}
This means that we allow the reference frame in which we compute
the trigonometric functions to be rotated by an angle $\alpha$.
One can check that they satisfy a commutation relation identical to
Eq.~(\ref{CLS}). Therefore, the associated uncertainty relations are
\begin{equation}
\label{uSC}
  ( \Delta S_{\alpha} )^2 ( \Delta L )^2  \ge
  \frac{1}{4} |\langle \op{C}_{\alpha} \rangle |^2 \, ,
  \quad
  ( \Delta C_{\alpha})^2 (\Delta L )^2 \ge
  \frac{1}{4} |\langle \op{S}_{\alpha} \rangle |^2 \, .
\end{equation}
Since Eqs.~(\ref{uSC}) are fully equivalent to Eq.~(\ref{disp}),
they cannot be saturated simultaneously. In fact, there are
further unfavorable aspects of them that have been reviewed in 
Ref.~\cite{Uffink:1990}. 

A common way of going on is to look for intelligent states
minimizing, e. g., the first one of these equations.  Although this
can be seen as dealing only with ``half'' the uncertainty principle,
the resulting states are often referred to as circular squeezed
states~\cite{Kowalski:2002} and exhibit amazing properties. They are
defined by
\begin{equation}
  \label{intelligent}
  ( \op{L} - i \kappa \op{C}_{\alpha} ) | \Psi \rangle = 
 \lambda |\Psi \rangle,
\end{equation}
where $\kappa$ and $\lambda $ are real parameters. Using the
angle representation, this extremal equation reads as
\begin{equation}
\label{von_mises}
    - i \frac{d}{d \phi } \Psi(\phi) =
    [  \lambda + i \kappa \cos(\phi + \alpha) ] \Psi(\phi) \, ,
\end{equation}
whose integration yields the normalized solution
\begin{equation}
\label{solution}
    \Psi (\phi) =   \frac{1}{\sqrt{2 \pi I_0(2 \kappa)}}
   \exp [ i \lambda \phi + \kappa \cos (\phi + \alpha ) ] \, ,
\end{equation}
$I_{0}$ being the modified Bessel function of order 0.  These are
called von Mises states, since the associated probability distribution
is precisely the von Mises, a very close analog of the Gaussian
distribution on the circle~\cite{Rehacek:2008ss}. The meaning of the
parameters is clear: $\lambda$ is the mean value of the angular
momentum, whereas $ \kappa $ determines the angular spread.

Next, we observe that the associated uncertainty relation in
Eq.~(\ref{uSC}) can be cast in the form
\begin{equation}
  \label{uncertainty1}
  (\Delta L)^2 \ge  U^2 \equiv  \frac{1}{4} \max_{\alpha}
  \frac{| \langle \op{C}_{\alpha} \rangle |^2}{(\Delta S_{\alpha})^2} \, .
\end{equation}
Let us introduce the following vectors
\begin{equation}
  \mathbf{x}  = \left (
    \begin{array}{c}
      \cos \alpha \\
      \sin \alpha \\
    \end{array}
  \right ) \, ,  
  \qquad
  \mathbf{c} = \left (
    \begin{array}{c}
      \langle C \rangle \\
      \langle S  \rangle \\
    \end{array}
  \right ) \, ,  
\end{equation}
and the covariance matrix
\begin{equation}
  \label{eq:covmat}
\mathbf{\Gamma}  = \left (
    \begin{array}{cc}
      (\Delta S)^2   & \Delta (SC) \\
      \Delta (CS) & (\Delta C)^2 \\
    \end{array}
  \right ),
\end{equation}
where $\Delta (C S) = \langle \op{C} \op{S} \rangle - \langle \op{C}
\rangle \langle \op{S} \rangle$. Then, $U^{2}$ can be written
as
\begin{equation}
  \label{performance2}
  U^2 = \frac{1}{4} \max_{ |\mathbf{x}| =1}
  \frac{( \mathbf{c}^{t} \, \mathbf{x}  )^2}
  {\mathbf{x}^{t} \, \mathbf{\Gamma} \, \mathbf{x}} \, ,
\end{equation}
and the superscript $t$ denotes the transpose. The optimization over
$\mathbf{x}$ can be easily performed, getting   
\begin{equation}
  \frac{\mathbf{c}^{t} \, \mathbf{x}}
  {\mathbf{x}^{t} \, \mathbf{\Gamma} \, \mathbf{x}}  \,
  \mathbf{c}  -
  \left ( \frac{\mathbf{c}^{t} \, \mathbf{x}}
  {\mathbf{x}^{t} \, \mathbf{\Gamma} \, \mathbf{x}} \right )^{2}
  \mathbf{\Gamma} \, \mathbf{x} = 0 \, ,
\end{equation}
whose solution gives the optimal value
\begin{equation}
\label{performance4}
    U^2 =  \frac{1}{4}  \mathbf{c}^{t} \mathbf{\Gamma}^{-1}  \mathbf{c} \, .
\end{equation}
We stress that while the variances $(\Delta C)^{2}$ and $(\Delta S)^{2}$
are not invariant under rotations of the state around the $z$ axis,
this is not the case with $U^{2}$, which constitutes a major
advantage. In addition,  $U^{2}$  combines the moments of $\op{C}$ and 
$\op{S}$ in  a rather nontrivial way, since
\begin{eqnarray}
 | \langle \op{E} \rangle |^2 & = &   
1 - \Tr \mathbf{\Gamma} \, ,  \nonumber \\
&  & \\ 
| \langle (\Delta E)^2 \rangle |^2 & = &  
( \Tr \mathbf{\Gamma})^{2}  - 4 \det  \mathbf{\Gamma} \, . 
\nonumber
\end{eqnarray}

The performance measure $U^{2}$ can be interpreted as a projection of
the noise into the direction of the preferred angle, analogously to
what was done for the ellipse representing a squeezed state in phase
space~\cite{Hradil:1991rm}.

Denoting by $\gamma_{-}$ and $\gamma_{+} $ the smaller and larger
eigenvalues of $\mathbf{\Gamma}$, a simple calculation allows us to
estimate 
\begin{equation}
  \label{estim}
  U^2   \ge  \frac{1}{4 \gamma_{+}} |\mathbf{c}|^{2} \equiv V^{2}
   \ge  
  \frac{1}{4} \frac{[ 1 - (\Delta E)^2]}{(\Delta E)^2} \, ,
\end{equation}
where we have introduced a new resolution performance
\begin{equation}
  V^2   = \frac{1}{4}
  \frac{2 (1 - \Tr \mathbf{\Gamma})}{\Tr \mathbf{\Gamma} +
   \sqrt{ (\Tr \mathbf{\Gamma} )^2 - 4 \det \mathbf{\Gamma}}} \, ,
\end{equation}
that combines the two basic invariants of $\mathbf{\Gamma}$. 
Notice that $V^{2}$ is related to the covariance matrix
(\ref{eq:covmat}) pretty much in the same way as the degree 
of polarization is linked to the polarization matrix. 
As we can see, it gives intermediate values between the bound in Eq.~(\ref{disp})  (which cannot be attained for nontrivial states) 
and the one in  Eq.~(\ref{uncertainty1}) (which is saturated by all the von
Mises states).

The second inequality in Eq.~(\ref{estim}) is saturated only in
trivial instances, such as, e. g., the eigenstates of
$\op{L}$~\cite{Rehacek:2008ss}. A condition for the first inequality
to be saturated is
\begin{equation}
  \label{eq:DCS}
  \Delta (CS) = 0 \, .
\end{equation}
This holds if the associated probability distribution is symmetrical
about some reference angle $\phi_0$, that is, $P (\phi_0 + \phi) = P
(\phi_0 - \phi)$. In addition, $U^2 = V^2$ also implies the additional
constraint
\begin{equation}
  \label{eq:condsim}
  (\Delta S)^{2} \ge (\Delta C)^{2} \, .
\end{equation}
The Von Mises states are among those satisfying  conditions
(\ref{eq:DCS}) and (\ref{eq:condsim}). For the other cases, one has
$U^{2} > V^{2}$.

\begin{figure}[t]
   \includegraphics[height=7cm]{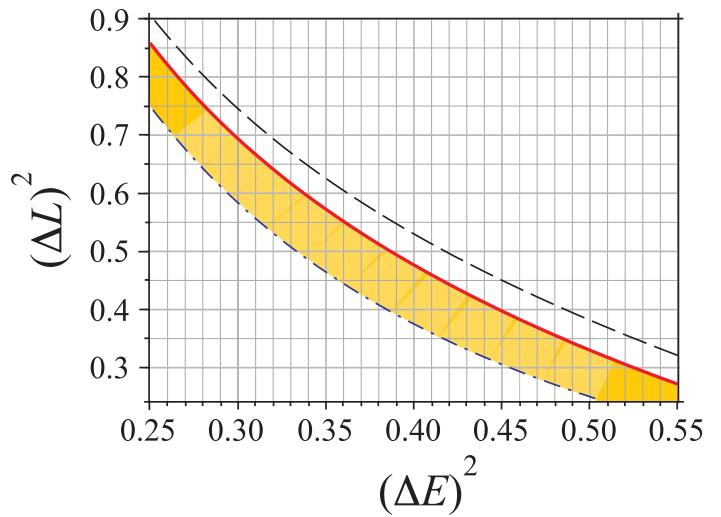}
  \caption{(Color online) Plot of the different bounds for $(\Delta
    L)^{2}$ in terms of the dispersion $\Delta E$ for the state
    (\ref{eq:catas}).  From bottom to top, we show the equality in
    Eq.~(\ref{disp}) (blue dashed-dotted line), in Eq.~(\ref{eq:uncV})
    (thick red line), and in Eq.~(\ref{uncertainty1}) (black dashed
    line).}
  \label{Fig1}
\end{figure}

In consequence, the inequality
\begin{equation}
  \label{eq:uncV}
  (\Delta L)^2 \ge  {V}^2
\end{equation}
is always true, significantly improves the standard bound in
Eq.~(\ref{disp}), and the right-hand side is saturable. Given that a
useful performance should be a simple expression of measurable quantities, 
we opt for using $V^2$, which depends on the two basic invariants of 
the covariance  matrix as one might expect, instead of $U^2$.  This latter 
quantity, in general, provides a slightly tighter bound.  However, 
as we have shown above, for the majority of states of interest the two bounds Eq.~(\ref{uncertainty1}) and (\ref{eq:uncV}) coincide, and both are saturated 
by the von Mises states.  The difference between $U^2$ and $V^2$
is in most cases unimportant and more than compensated by the utility
and feasibility of the proposed uncertainty relation Eq.~(\ref{eq:uncV}).

In Fig.~\ref{Fig1} we have condensed all this information for the
state
\begin{equation}
  \label{eq:catas}
  \Psi (\phi) = \frac{1}{\sqrt{4 \pi I_{0} (2 \kappa)}}
\left [ \exp (\kappa \cos \phi) - i \exp (i \phi +
\kappa \cos \phi ) \right ] \, ,
\end{equation}
which corresponds to the superposition of two von Mises states
with $\langle \op{L} \rangle = 0$ and  $\langle \op{L} \rangle = 1$.
This can be seen as an angular counterpart of a cat state, with a 
probability distribution
\begin{equation}
  \label{eq:disca}
  P (\phi ) = (1 + \sin \phi ) \exp (- 2 \kappa \cos \phi) \, ,
\end{equation}
displaying a lack of symmetry.  The proposed bound
(\ref{eq:uncV}) constitutes a good improvement over the standard
one (\ref{disp}), as we can see in the figure: all the area shaded 
corresponds to the values of $(\Delta L)^2$ that, for a given
angular fluctuation $(\Delta E)^2$, are permitted by the standard
uncertainty relation but not allowed according to our proposal.
Obviously, the strongest bound in Fig.~\ref{Fig1} is provided by
relation (\ref{uncertainty1}).  The family of states
(\ref{eq:disca}) was deliberately chosen so as to make the difference
between $V^2$ and $U^2$ large; but even in that case the improvement of
both (\ref{eq:uncV}) and (\ref{uncertainty1}) over the standard bound
is seen to be much larger than the difference between them.

Our arguments support the role of von Mises distribution on the circle
as an analog of the Gaussian distribution on the line, at least as far
as the uncertainty product is concerned. However, things may be not
that simple with other aspects of quantum behavior. Indeed, our latest 
research indicates that the Wigner function of von Mises states is 
not positive (as it happens for Gaussian states in the line), since 
this property is reserved exclusively to the angular momentum
eigenstates~\cite{Rigas:2008ac,Rigas:2009rt}.

Finally, we observe that one could introduce a ladder
operator~\cite{Kowalski:1996}
\begin{equation}
  \label{eq:defW}
  \hat{X} =  e^{-\hat{L} - 1/2} \, \hat{E} \, .
\end{equation}
Since this can be expressed also in terms of the non-unitary
transformation $\op{X} = e^{ \op{L}^2/2} \op{E}
e^{- \op{L}^2/2}$~\cite{Kastrup:2006wh}, the commutator $ [\op{X},
\op{L}] = \op{X}$ still remains valid.  The construction of the
accessible lower bound in this paper can be thus repeated, provided
the role of $\op{C}$ and $\op{S}$ is now taken by the quadrature-like
operators
\begin{equation}
  \op{Q} = \frac{1}{2} (\op{X} + \op{X}^{\dagger}) \, , 
  \qquad
  \op{P} = \frac{1}{2i} (\op{X} - \op{X}^{\dagger}) \, . 
\end{equation}
Obviously, the (unnormalized) extremal states for these operators are
given by the von Mises states, but transformed by $ e^{-\op{L}^2/2}$.

In summary, what we expect to have accomplished here is to present
convincing arguments for the use of a new angular resolution measure
that involves only invariant and measurable quantities, has
no problem with periodicity, and gives an improved feasible criterion
to assess minimal angle fluctuations.

We acknowledge discussions with Hubert de Guise.  This work was
supported by the Research Project of the Czech Ministry of Education
``Measurement and Information in Optics'' MSM 6198959213 and the Spanish
Research Directorate, Grant No FIS2008-04356.


\end{document}